\documentstyle[epsf,12pt]{article}
\newlength{\mytopmargin}
\newlength{\myleftmargin}
\def\zz{\rlx\hbox{\small \sf Z\kern-.4em Z}}

\newcommand{\ml}{\langle}
\newcommand{\mg}{\rangle}

\begin{document}

\vspace{1cm}
\noindent
\begin{center}{   \large \bf
Exact statistical properties of the zeros of \\ complex
random polynomials}  
\end{center}
\vspace{5mm}

\noindent
\begin{center} 
 P.J.~Forrester and G.~Honner \\

\it Department of Mathematics and Statistics, \\
University of Melbourne, Parkville, Victoria
3052, Australia
\end{center}
\vspace{.5cm}

\small
\begin{quote}
The zeros of complex Gaussian random polynomials, with coefficients such that
the density in the underlying complex space is uniform, are known to
have the same statistical properties as the zeros of the coherent state
representation of one-dimensional chaotic quantum systems. We extend the
interpretation of these polynomials by showing that they also arise
 as the wave function for a quantum particle in a magnetic field
constructed from a random superposition
of states in the lowest Landau level. A study of the statistical properties
of the zeros is undertaken using exact formulas for the one and two
point distribution functions. Attention is focussed on the moments of the
two-point correlation in the bulk, the variance of a linear statistic,
and the asymptotic form of the two-point correlation at the boundary.
A comparison is made with  the same quantities for the eigenvalues
of complex Gaussian random matrices.
\end{quote}

\vspace{.5cm}
\noindent
\section{Introduction}
\setcounter{equation}{0}
The zeros of random polynomials have received recent attention as providing
the universality class for the zeros of the coherent state representation
of one-dimensional chaotic quantum systems \cite{1}-\cite{6}. 
Moreover, denoting the
coefficients by $\alpha_j$ so the polynomial reads
\begin{equation}\label{1.1}
p(z) = \sum_{j=0}^N \alpha_j z^j,
\end{equation}
it has recently been realized that in the particular case that the real and
imaginary parts of $\alpha_j$ are independent Gaussian random variables with
mean zero and standard deviation $\sigma_j$, which we write as
\begin{equation}\label{1.2}
\alpha_j = {\rm N}[0,\sigma_j] + i {\rm N}[0,\sigma_j],
\end{equation}
the zero distribution is a solvable model \cite{7} - \cite{14}. 
Thus the probability
density function (p.d.f.) can be computed in terms of the $\sigma_j$
\cite{7}, as can the $n$-particle distribution functions for each
$n =1,2,\dots$ \cite{9}. 

Our study is motivated by the explicit form of
the zeros p.d.f.,
\begin{equation}\label{1.3}
p(z_1,\dots,z_N) = \pi N! \bigg (\prod_{l=0}^N {1 \over \pi \sigma_l^2}
\bigg )
{\prod_{1 \le j < k \le N} |z_k - z_j|^2 \over
(\sum_{j=0}^N |e_j|^2/\sigma_{N-j}^2)^N},
\end{equation}
where 
\begin{equation}\label{1.4}
e_j := \sum_{1 \le p_1 < \cdots <p_j \le N} z_{p_1} \cdots z_{p_j}
\end{equation}
and $e_0 := 1$. As pointed out in \cite{9}, the factor
\begin{equation}\label{1.5}
\prod_{1 \le j < k \le N} |z_k - z_j|^2  = e^{-2 \sum_{j < k}
\log |z_k - z_j|}
\end{equation}
can be interpreted as the Boltzmann factor for a two-dimensional classical
$N$ particle system with pairwise logarithmic repulsion at inverse temperature
$\beta = 2$. The particles are prevented from repelling to infinity by an 
extensive many body interaction with Boltzmann factor
$(\sum_{j=0}^N |e_j|^2/\sigma_{N-j}^2)^{-N}$. Furthermore, with the particular
standard deviation
\begin{equation}\label{1.5a}
\sigma_j = {1 \over (j!)^{1/2}}
\end{equation}
the density of zeros is, to leading order, uniform inside a disk of radius
$\sqrt{N}$ and zero outside this disk \cite{5} (see also Section 5).
This latter feature is shared by the
eigenvalue p.d.f.~for complex
random matrices with each element independent and chosen from the
complex Gaussian distribution N$[0,1/\sqrt{2}]+ i {\rm N}[0,1/\sqrt{2}]$,
which is proportional to
\begin{equation}\label{1.6}
\prod_{j=1}^N e^{-\sum_{j=1}^N |z_j|^2} 
\prod_{1 \le j < k \le N} |z_k - z_j|^2.
\end{equation}

Let us consider further the eigenvalue p.d.f.~(\ref{1.6}).
Some insight into the behaviour of the
density  can be obtained by noting that for
$|z| < \sqrt{N}$, (\ref{1.6}) is proportional to the  Boltzmann factor
of the two-dimensional one-component plasma (2dOCP) at the coupling $\beta =2$.
The factor $\exp (-\sum_{j=1}^N |z_j|^2)$ there results
from the potential energy
between a uniform background of charge density $-1/\pi$ and $N$ unit
charges interacting via the logarithmic potential, so the fact that the
density has support in a disk of radius $\sqrt{N}$ can be
interpreted as the system having a preference for charge  neutrality.
Moreover, the plasma system is an example of a two-dimensional Coulomb
system in its conductive phase, and so exhibits universal properties
for the behaviour of the two-point correlation, both in the bulk
and at the boundary $|z| \approx \sqrt{N}$ (see e.g.~\cite{15,16}).
 These special forms
in turn imply a special form for the variance of a linear statistic
\cite{17}. The eigenvalue distribution of course must also
exhibit these Coulomb properties. Because the zeros p.d.f.~has
the factor (\ref{1.5}) in common with (\ref{1.6}), and with the
choice (\ref{1.5a}) has the density in common too, we are led to
investigate for the zeros distribution
the two-point correlation in the bulk and at the
boundary  $|z| \approx \sqrt{N}$, and the  variance of a linear statistic,
to see if they exhibit universal Coulomb like properties.

The plan of the paper is as follows. In Section 2 an interpretation of
the complex Gaussian random polynomial with variance (\ref{1.5a}) will be
given as a random wave function associated with a superposition of
lowest Landau states in the plane. A similar interpretation will be
given to the complex Gaussian random polynomial with variance of the
coefficients
\begin{equation}\label{1.7a}
\sigma^2_j = {N \choose j},
\end{equation}
considered originally in \cite{9},
except this time the Landau states are those for a particle confined to
the surface of the sphere with a perpendicular magnetic field.
In Section 3 the known \cite{9} general formulas for the one and
two-point distributions are revised, and these are specialized to the
cases (\ref{1.5a}) and (\ref{1.7a}). The investigation proper is carried
out in Sections 4 and 5. In Section 4 we study bulk properties of the
two-point correlation function, and show how the formulas obtained for
its moments have consequence with regard to fluctuation formulas for
the number of particles in a region $\Lambda$ as $|\Lambda| \to 0$,
and the variance of a linear statistic as a function of $N$ in the
sphere case (\ref{1.7a}). The behaviour of the density and
two-point correlation at the boundary of support in the disk
case (\ref{1.5a}) is the subject of Section 5. The paper ends with a
brief summary of our results.

\section{Quantum mechanical interpretation}
\setcounter{equation}{0}
A number of studies \cite{1}-\cite{6} have related complex polynomials to the
wave function of chaotic quantum systems. More precisely, complex random
polynomials with variances (\ref{1.5a}) and (\ref{1.7a}) arise as the
Bargmann or coherent state representation of one-dimensional chaotic
wave functions and chaotic spin states respectively. In this section
we point out that these complex random polynomials also occur as random
superpositions of lowest Landau level states for a charged quantum
particle, in a plane and on the surface of a sphere, in the presence
of a perpendicular magnetic field.

Consider first the planar problem. Let $B$ denote the strength of the
magnetic field, and suppose the vector potential $\vec{A}$ is chosen in
the symmetric gauge so that
$$
\vec{A} = {B \over 2}( - y \hat{x} + x \hat{y} ).
$$
Then it is well known (see e.g.~\cite{18}) that the lowest Landau level
(i.e.~ground state) consists of any wave function of the form
\begin{equation}\label{2.f}
e^{-|z|^2/4\ell^2} f(\bar{z}),
\end{equation}
where $z:= x + iy$, $\ell := \sqrt{\hbar c / eB}$ ($e$ denotes the charge of the
particles) is the magnetic length
and $f$ is analytic in $\bar{z}$. Furthermore the states
(\ref{2.f}) can be separated by requiring that they be simultaneous
eigenstates of the quantum mechanical operator corresponding to the
square of the centre of the cyclotron orbit. This operator has
eigenvalues $R_n^2 = (2n+1)\ell^2$ ($n=0,1,\dots$) with corresponding
normalized eigenfunctions of the form (\ref{2.f}) given by
\begin{equation}\label{2.f1}
\psi_n(\vec{r}) = {\bar{z}^n e^{-|z|^2/2} \over \pi^{1/2} (n!)^{1/2}}
\end{equation}
(here the units have been chosen so that $2l^2 = 1$). A state
$\phi(\vec{r})$ consisting of the first $N+1$ of these states has the form
\begin{equation}\label{2.r}
\phi(\vec{r}) = {1 \over {\cal N} \pi^{1/2}} e^{-|z|^2/2} p(\bar{z}),
\qquad p(w) := \sum_{n=0}^N {\alpha_n \over \sqrt{n!}} w^n
\end{equation}
with ${\cal N} := (\sum_{j=0}^N |\alpha_n|^2)^{1/2}$. Choosing the
state to be random by selecting each coefficient $\alpha_n$ at
random from the complex Gaussian distribution ${\rm N}[0,1] + 
i {\rm N}[0,1]$,
we see immediately that $p(w)$ is a complex Gaussian random polynomial
with variance (\ref{1.5a}).

For the sphere problem, the magnetic field is due to a monopole, which
must satisy \cite{19} the quantization condition $B=N\hbar c/2 e R^2$,
$N=1,2,\dots$. Furthermore the Hamiltonian operator $H$ can be written
\cite{18.5}
\begin{equation}\label{2.H}
H = {1 \over 2 m R^2} \Big ( \vec{L}^2 - {\hbar^2 N^2 \over 4} \Big ),
\end{equation}
where the components of $L$ obey the canonical commutation relations
for angular momentum
$$
[L_x,L_y] = i \hbar L_z, \quad
[L_y,L_z] = i \hbar L_x, \quad
[L_z,L_x] = i \hbar L_y.
$$
Hence the allowed values of $\vec{L}^2$ are $l(l+1)\hbar^2$,
$l = 0,{1 \over 2}, 1, {3 \over 2},\dots$ and so for (\ref{2.H}) to
be positive definite the smallest allowed value of $l$ is $N/2$,
which thus corresponds to the lowest Landau level. The states in this
level can be distinguished by seeking simultaneous eigenfunctions
of $L_z$ and $H$. This shows that there is a $N+1$ fold degeneracy,
which is spanned by the orthogonal functions
$$
\psi_m(\theta,\phi) = \Big \{
{N/2 + 1 \over 4 \pi R^2} {N \choose N/2 + m} \Big \}^{1/2}
u^{N/2 + m} v^{N/2 - m} e^{-iN\phi / 2}, \quad
m=-N/2,-N/2+1,\dots,N/2,
$$
where 
$$
u := \cos {\theta \over 2} \, e^{i \phi / 2}, \qquad
v = - i \sin {\theta \over 2} \,  e^{-i \phi / 2}
$$
denote the Cayley-Klein parameters and $(\theta,\phi)$ are the usual
spherical coordinates. Forming a linear combination of such states gives
the state
$$
\Phi(\theta,\phi) = {1 \over {\cal N}}
\Big ( {N/2 + 1 \over 4 \pi R^2} \Big )^{1/2}
(- i \sin \theta/2)^N e^{- i N \phi /2} p ( e^{i \phi} \cot {\theta \over 2} ),
\quad p(z) = \sum_{n=0}^N {N \choose n}^{1/2} \alpha_n z^n.
$$
If this state is chosen at random by specifying the $\alpha_n$ as belonging
to the random complex Gaussian distribution ${\rm N}[0,1]+ i {\rm N}[0,1]$
as in (\ref{2.r}), then the polynomial $p(z)$
is a complex random polynomial with variance (\ref{1.7a}).
Furthermore, the mapping 
\begin{equation}\label{2.proj}
z = e^{i \phi} \cot {\theta \over 2}
\end{equation}
 represents
a stereographic projection from the north pole of the unit sphere to
a plane passing throught its equator, showing that $p$ as a function
the spherical coordinates $(\theta,\phi)$ has $N$ zeros on the surface of
the unit sphere. Such a factorization of a spin state is due to
Majorana \cite{20}.

\section{The distribution functions}
As mentioned in the Introduction, the general $n$-point distribution
function $\rho_{(n)}$ for the zeros of the random polynomial (\ref{1.1}) can
be computed exactly in the case that the coefficients are complex
Gaussian random variables drawn from the distribution (\ref{1.2})
\cite{9}. Our interest is in this explicit form in the cases
$n=1$ and $n=2$. 

The case $n=1$ corresponds to the density of zeros.
This one point distribution function has the explicit form
\cite{9,14}
\begin{equation}\label{3.den}
\pi \rho_{(1)}(\vec{r}) = {\partial^2 \over \partial z \partial \bar{z}}
\log \ml p(z) p(\bar{z}) \mg,
\end{equation}
where the averages are with respect to the distribution of the
coefficients. Here $\vec{r} := (x,y)$
with $x + i y := z$. Performing this average in the case that the
distribution of the coefficients $\alpha_j$ is given by (\ref{1.5a})
gives
\begin{equation}\label{3.den1}
\pi \rho_{(1)}(\vec{r}) = {\partial^2 \over \partial z \partial \bar{z}}
\log e(|z|^2;N) 
\end{equation}
where
\begin{equation}\label{3.den2}
e(u;n) := \sum_{p=0}^n {u^p \over p!}.
\end{equation}
Notice that with $z$ fixed $e(|z|^2;N) \to e^{|z|^2}$ as $N \to \infty$,
so the formula (\ref{3.den1}) gives that
\begin{equation}\label{3.den1'}
\rho_{(1)}(\vec{r}) = {1 \over \pi}
\end{equation}
in the bulk.

Repeating the calculation which led to (\ref{3.den1})
with the coefficients given by (\ref{1.7a})
shows that the corresponding density is
\begin{equation}\label{3.den3}
\pi \rho_{(1)}(\vec{r}) = {\partial^2 \over \partial z \partial \bar{z}}
\log \Big ( 1 + |z|^2 \Big )^N.
\end{equation}
Due to the physical origin of the variances (\ref{1.5a}), it is
natural to
perform a stereographic projection onto the sphere according to the
mapping (\ref{2.proj}) so that
\begin{equation}\label{3.cv}
\rho_{(1)}(\vec{r}) dx dy = \rho_{(1)}(\hat{s}) {4 \,dS \over (1 + |z|^2)^2},
\end{equation}
where $\rho(\hat{s})$ refers to the density on the unit sphere 
with surface element $dS$ at the point $\hat{s}$ determined by (\ref{2.proj})
and
$4/(1 + |z|^2)^2$ is the Jacobian for the change of variables.
This gives \cite{8,9,14}
\begin{equation}\label{3.sd1}
\rho_{(1)}(\hat{s}) = {N \over 4 \pi},
\end{equation}
thus showing that the density is uniform.

Not surprisingly, the complexity of the explicit form of the 
$n$-point distribution increases with $n$. We will not go beyond the
$n=2$ case, for which the distribution has the explicit form
\begin{equation}\label{3.den4}
\pi^2 \rho_{(2)}(\vec{r}_1,\vec{r}_2) =
{{\rm per} ( C - B^\dagger A^{-1} B ) \over \det A}.
\end{equation}
The notation per denotes the permanent, which for a $2 \times 2$ matrix is
defined by
\begin{equation}\label{3.den5}
{\rm per} \left [ \begin{array}{cc} a & b \\ c & d \end{array} \right ] :=
a d + bc,
\end{equation}
and with $p_j := p(z_j)$, $p_j' := p'(z_j)$, $(j=1,2$ and the dash
denotes differentiation, the matrices $A$,
$B$ and $C$ are defined by
\begin{equation}\label{3.den6}
A := \left [ \begin{array}{cc}  \ml p_1 \bar{p}_1 \mg & \ml p_1\bar{p}_2 \mg \\
\ml p_2 \bar{p}_1 \mg & \ml p_2 \bar{p}_2 \mg \end{array}\right ], \quad
B := \left [ \begin{array}{cc}  \ml p_1 \bar{p}_1' \mg & \ml p_1\bar{p}_2' \mg\\
\ml p_2 \bar{p}_1' \mg & \ml p_2 \bar{p}_2'\mg \end{array} \right ], \quad
C := \left [ \begin{array}{cc}  \ml p_1' \bar{p}_1' \mg & \ml p_1'\bar{p}_2' \mg\\
\ml p_2' \bar{p}_1' \mg & \ml p_2' \bar{p}_2'\mg \end{array} \right ].
\end{equation}
Writing 
$$
A^{-1} = {1 \over \det A} 
\left [ \begin{array}{cc} a_{22} & -a_{12} \\ -a_{21} & a_{11} \end{array} \right ]
=: {1 \over \det A} A^D
$$
we see that (\ref{3.den4}) can be written in the equivalent form
\begin{equation}\label{3.den4'}
\pi^2 \rho_{(2)}(\vec{r}_1,\vec{r}_2) =
{{\rm per} ( dC - B^\dagger A^{D} B ) \over d^3}, \quad d = {\rm det} A.
\end{equation}

Consider now the evaluation of the matrix elements in (\ref{3.den6})
in the case of (\ref{1.5a}). These are made explicit by referring to
the formulas
\begin{eqnarray}\label{3.den6'}
\ml p_i \bar{p}_j \mg = e(z_i \bar{z}_j; N) &
\ml p_i' \bar{p}_j \mg = \bar{z}_j e(z_i \bar{z}_j; N-1) \nonumber \\
\ml p_i \bar{p}_j' \mg = {z}_i e(z_i \bar{z}_j; N-1) &
\ml p_i' \bar{p}_j' \mg = z_i \bar{z}_j e(z_i \bar{z}_j;N-2) +
e(z_i \bar{z}_j; N-1),
\end{eqnarray}
where $e(u;n)$ is defined by (\ref{3.den2}), which in turn follow from the
definitions. At this stage we will not attempt to further develop
(\ref{3.den4'}); this task will be undertaken in Section 5 when the
$N \to \infty$ limit in the neighbourhood of the boundary $|z| =
\sqrt{N}$ is computed. However we will note the previously
computed \cite{9} $N \to \infty$ limiting value with $\vec{r}_1$ and
$\vec{r}_2$ fixed. This limit gives the 2-point distribution
function in the bulk, and has the explicit evaluation\footnote{
In \cite{9} the factor of 1/2 in the argument of
$h(|\vec{r}_1 - \vec{r}_2|^2/2)$ is missing; this has been subsequently
corrected in \cite{6}.}
\begin{equation}\label{3.bt}
 \rho_{(2)}(\vec{r}_1,\vec{r}_2) = {1 \over \pi^2}
h(|\vec{r}_1 - \vec{r}_2|^2/2),
\qquad h(x) := {(\sinh^2 x + x^2) \cosh x - 2 x \sinh x \over
\sinh^3 x}.
\end{equation}

In the case of the variances (\ref{1.7a}) we also require knowledge of
the matrix elements in (\ref{3.den6}). A simple calculation gives \cite{9}
\begin{eqnarray}\label{3.sub}
\ml p_i \bar{p}_j \mg = (1 + z_i \bar{z}_j)^N &
\ml p_i' \bar{p}_j \mg = N \bar{z}_j (1 + z_i\bar{z}_j)^{N-1}  \nonumber \\
\ml p_i \bar{p}_j' \mg = N z_i (1 + z_i\bar{z}_j)^{N-1} &
\ml p_i' \bar{p}_j' \mg = N(1 + N z_i \bar{z}_j) (1 + z_i \bar{z}_j)^{N-2}.
\end{eqnarray}
In Section 4 we will have application for the corresponding evaluation
of (\ref{3.den4'}) in the finite system, when projected onto the
sphere according to the transformation (\ref{2.proj}). By substituting
(\ref{3.sub}) into (\ref{3.den6}), and then the result into (\ref{3.den4'}), 
expanding the permanent according to (\ref{3.den5})
and performing some elementary but tedious manipulations we find
\begin{eqnarray}\label{3.sp}
\rho_{(2)}^T(\hat{s}_1,\hat{s}_2) & := & \rho_{(2)}(\hat{s}_1,\hat{s}_2) -
\rho_{(1)}(\hat{s}_1) \rho_{(1)}(\hat{s}_2) \nonumber \\
& = & {N^2 \over 16 \pi^2}
{f^{2N-4} \over 1 - f^{2N}} \Big (
1 + f^4 - 2N {1 - f^4 \over 1 - f^{2N}} +
N^2 {(1+f^{2N})(1 - f^2)^2 \over (1 - f^{2N})^2 } \Big )
\end{eqnarray}
where
\begin{equation}\label{3.20'}
f^2 = (f(\hat{s}_1,\hat{s}_2))^2 := {1 \over 2}(1 + \hat{s}_1 \cdot \hat{s}_2).
\end{equation}

\section{The distribution functions in the bulk and fluctuation formulas}
\setcounter{equation}{0}
\subsection{The two-point function in the bulk}
In this section properties of the bulk two-point distribution function
for the 2dOCP at $\beta = 2$, or equivalently for the  eigenvalues of
complex random matrices specified in the Introduction, will be recalled
and contrasted with the corresponding properties of the bulk 
two-point distribution function (\ref{3.bt}). Now, for a bulk density
$\rho_{(1)}(\vec{r}) = 1/ \pi$, the bulk two-point distribution function
for the 2dOCP at $\beta = 2$ has the explicit form \cite{21,22}
\begin{equation}\label{3.bt1}
\rho_{{\rm OCP}\,(2)}(\vec{r}_1,\vec{r}_2) = {1 \over \pi^2 }
\Big ( 1 - e^{-|\vec{r}_1 - \vec{r}_2|^2} \Big ),
\end{equation}
or equivalently the truncated distribution has the explicit form
\begin{equation}\label{3.bt2}
\rho_{{\rm OCP}\,(2)}^{T}(\vec{r}_1,\vec{r}_2) = - {1 \over \pi^2}
e^{-|\vec{r}_1 - \vec{r}_2|^2}.
\end{equation}

One feature of (\ref{3.bt1}) is that it is a strictly
negative monotonic function and is its own large separation
asymptotic form,
$$
\rho_{{\rm OCP}\,(2)}^{T}(\vec{r},\vec{0})
\mathop{\sim}\limits_{|\vec{r}| \to \infty} - {1 \over \pi^2}
e^{-|\vec{r}|^2}.
$$
In contrast (\ref{3.bt}) reaches a maximum before approaching asymptotically
approaching zero from above (see e.g.~the plot in \cite{9}).
The large separation asymptotic form is computed from (\ref{3.bt}) to be
$$
\rho_{(2)}^{T}(\vec{r},\vec{0})
\mathop{\sim}\limits_{|\vec{r}| \to \infty} |\vec{r}|^4
e^{-|\vec{r}|^2}.
$$

The zeroth and second moments of  $\rho_{{\rm OCP}\,(2)}^{T}$ exhibit
particular universal properties of Coulomb systems.
The zeroth moment is evaluated as
$$
\int_{{\rm R}^2} \rho_{{\rm OCP}\,(2)}^{T}(\vec{r},\vec{0}) \, d\vec{r} =
- {1 \over \pi},
$$
which can be written in the equivalent form
\begin{equation}\label{3.csr}
\int_{{\rm R}^2} \Big ( \rho_{{\rm OCP}\,(2)}^{T}(\vec{r},\vec{0}) 
+ \rho_{(1)}(\vec{r}) \delta(\vec{r}) \Big )
\, d\vec{r} = 0.
\end{equation}
In this guise the zeroth moment formula can be interpreted as an example
of a charge sum rule for Coulomb systems (see e.g.~\cite{15}) which says
the total charge of a fixed charge (here a charge at the origin) and its
screening cloud (represented by $\rho_{{\rm OCP}\,(2)}^{T}(\vec{r},\vec{0})$)
is zero.

The second moment of (\ref{3.bt1}) is evaluated as
\begin{equation}\label{3.ef}
\int_{{\rm R}^2} 
|\vec{r}|^2 \rho_{{\rm OCP}\,(2)}^{T}(\vec{r},\vec{0}) \, d\vec{r}
= - {1 \over \pi}.
\end{equation}
By invoking a linear response relation (see e.g.~\cite{15,16}) this
can be shown to be equivalent to the statement that the system perfectly
screens an external charge density in the long wavelength limit.
Thus suppose the 2dOCP for general $\beta$ is perturbed by an external
charge density $\epsilon e^{i\vec{k}\cdot \vec{r}}$. The potential
energy of the system is then changed by an amount 
\begin{eqnarray}
\delta U & := & - \epsilon \int_{{\bf R}^2}d\vec{r}\,' \,
\int_{{\bf R}^2}d\vec{r} \, \log|\vec{r}-\vec{r}\,'| 
n_{(1)}(\vec{r}\,') e^{i\vec{k}\cdot \vec{r}} \nonumber \\
& = & {2 \epsilon \pi \over |\vec{k}|^2} \tilde{n}_{(1)}(\vec{k}),
\end{eqnarray}
where $n_{(1)}(\vec{r})$ represents the microscopic density at the
point $\vec{r}$, $\tilde{n}_{(1)}(\vec{k})$ its two-dimensional
Fourier transform, and to obtain the second equality the result
\begin{equation}\label{3.ft}
- \int_{{\bf R}^2}d\vec{r} \, \log|\vec{r}| = {2 \pi \over |\vec{k}|^2}
\end{equation}
from the theory of generalized functions has been used.
Suppose now that the microscopic charge density at the point $\vec{r}$, 
which for the OCP is just $n_{(1)}(\vec{r})$, is observed. The linear
response relation gives that
\begin{equation}\label{3.lr}
\ml n_{(1)}(\vec{r}) \mg_\epsilon - \ml n_{(1)}(\vec{r}) \mg_0
= - \beta \ml n_{(1)}(\vec{r}) \delta U \mg_0^T,
\end{equation}
where the subscript $(\epsilon)$ indicates the system in the presence of
the perturbation. Now, a characteristic of a Coulomb system in its conductive
phase is that it will perfectly screen an external charge density in the
long wavelength limit. This means that
$$
\ml n_{(1)}(\vec{r}) \mg_\epsilon - \ml n_{(1)}(\vec{r}) \mg_0
\mathop{\sim}\limits_{|\vec{k}| \to 0} - \epsilon e^{i\vec{k}\cdot \vec{r}}.
$$
Substituting this in (\ref{3.lr}), and using the translational invariance
of the system in the bulk, we see that
\begin{equation}\label{3.lr1}
\tilde{S}(\vec{k}) \mathop{\sim}\limits_{|\vec{k}| \to 0}
{|\vec{k}|^2 \over 2 \pi \beta}, \qquad
S(\vec{r}) := \rho_{(2)}^T(\vec{r},\vec{0}) + \rho \delta(\vec{r})
\end{equation}
This implies the charge sum formula (\ref{3.csr}) as well as the
second moment (Stillinger-Lovett) sum rule 
\begin{equation}\label{3.lr2}
\int_{{\bf R}^2} d\vec{r} \, |\vec{r}|^2 \rho_{{\rm OCP} \,(2)}^{
T}(\vec{r},\vec{0})
= -{2 \over \beta \pi},
\end{equation}
which reduces to (\ref{3.ef}) at $\beta = 2$.

Let us now compute the zeroth and second moments of (\ref{3.bt}), and
compare their values with those for (\ref{3.bt1}). Now, whereas the
computation of the moments in the case of (\ref{3.bt1}) is elementary,
it is not so simple to do likewise for (\ref{3.bt}). Fortunately the
function $h(x)$ in (\ref{3.bt}) can be written as the second
derivative of an elementary function. Thus by explicit calculation we
can check that
\begin{equation}\label{3.h}
h(x) = {1 \over 2} {d^2 \over dx^2} x^2 \coth x,
\end{equation}
or equivalently
\begin{equation}\label{3.h1}
h(x) - 1 = {1 \over 2} {d^2 \over dx^2} \Big (x^2 (\coth x - 1) \Big ).
\end{equation}
Making use of (\ref{3.h1}) we find that again the charge sum rule
(\ref{3.csr}) is satisfied, whereas for the second moment we find
\begin{equation}\label{3.h2}
\int_{{\bf R}^2} d\vec{r} \, |\vec{r}|^2 \rho_{(2)}^{
T}(\vec{r},\vec{0})
= 0.
\end{equation}
Although we have know that 
when viewed as a classical particle system the 
 p.d.f.~(\ref{1.3}) for the 
zeros distribution contains many body
forces in addition to the pair potential term (\ref{1.5}), it is of some
interest to compare the results (\ref{3.csr}) and (\ref{3.h2}) to
those for a two-dimensional classical fluid with purely pairwise interactions.
For long-range potentials with Fourier transform
$\tilde{v}(\vec{k}) \sim c |\vec{k}|^{-\alpha}$, $\alpha > 0$,
we can generalize the perfect screening argument which led to
(\ref{3.lr1}), this time obtaining
\begin{equation}\label{3.h3}
\tilde{S}(\vec{k}) \mathop{\sim}\limits_{|\vec{k}| \to 0}
{|\vec{k}|^\alpha \over c \beta}.
\end{equation}
Setting $\vec{k} = \vec{0}$ gives (\ref{3.csr}) in all cases.
However, as is well known (see e.g.~\cite{23.5}), 
it is only for even values of
$\alpha$ that (\ref{3.h3}) is compatible with a fast decay of $S(\vec{r})$.
In the Coulomb case $\alpha = 2$, (\ref{3.h3}) then gives the second
moment condition (\ref{3.lr2}). The next smallest even value is
$\alpha = 4$. Then (\ref{3.h3}) indeed implies that the second moment
vanishes, just as observed in (\ref{3.h2}) for the complex zeros.
However a pair potential in two-dimensions for which the Fourier
transform behaves as
$|\vec{k}|^4$ for $|\vec{k}| \to 0$, has the large $|\vec{r}|$ behaviour
in real space proportional to $-|\vec{r}|^2 \log |\vec{r}|$.
It is unlikely that a classical system with such strong repulsion can
be thermodynamically stable, even with a neutralizing background.

\subsection{Fluctuation formulas in the infinite system}
In general the properties of the two-point function of a many particle
system, in particular the value of its moments, have consequences regarding
fluctuation formulas for linear statistics. We recall that a linear
statistic is any observable (function) of the form
$A = \sum_{j=1}^N a(\vec{r_j})$. It is easy to see from the definitions
that the two-point function $S(\vec{r})$ (\ref{3.lr1})
is directly related to the
fluctuation (variance) of a linear statistic via the formula
\begin{eqnarray}\label{3.for}
{\rm Var}(A) & := & \ml A^2 \mg - \ml A \mg^2 \nonumber \\
& = &\int d\vec{r}_1 \, a(\vec{r}_1) \int d\vec{r}_2 \,  a(\vec{r}_2)
S(\vec{r}).
\end{eqnarray}

The linear statistic specified by $a(\vec{r}) = \chi_{\Lambda}(\vec{r})$
where $\chi_{\Lambda}$ denotes the indicator function of the domain
$\Lambda$: $\chi_{\Lambda}(\vec{r}) = 1$ if $\vec{r} \in \Lambda$,
$\chi_{\Lambda}(\vec{r}) = 0$ otherwise, measures the number of particles
in the region $\Lambda$. In this case Var$(A)$ is the number variance,
to be denoted $\Sigma_2(\Lambda)$ which specified the fluctuation from the
mean of the number of particles in $\Lambda$. It is a well known fact 
\cite{23.5} that whenever $\Lambda$ can be generated by a dilation from a
fixed domain $\Lambda_0$, and  the charge sum rule (\ref{3.csr}) holds,
the number variance has the asymptotic behaviour
\begin{equation}\label{3.asb}
\lim_{|\Lambda| \to \infty} {\Sigma_2(\Lambda) \over |\partial \Lambda|}
= - {1 \over \pi} \int_{{\bf R}^2} d\vec{r} \, |\vec{r}|
\rho_{(2)}^T(\vec{r},\vec{0}),
\end{equation}
assuming the integral exists.
Here, with $\Lambda$ a two-dimensional domain, $|\partial \Lambda|$ denotes
the length of the perimeter of $\Lambda$. Thus the fluctuations are
strongly suppressed in comparison to a compressible fluid for which
$\Sigma_2(\Lambda)$ is proportional to $|\Lambda|$.

In the case of the complex zeros the two-point function is specified
by (\ref{3.bt}). Making use of the formula (\ref{3.h1}) the first
moment in (\ref{3.asb}) can be computed exactly. For this purpose,
after introducing polar coordinates, we integrate by parts twice
and make use of the definite integral
$$
\int_0^\infty{t^{1/2} e^{-t} \over 1 - e^{-t}} dt =
{\sqrt{\pi} \over 2} \zeta ({3 \over 2}).
$$
Substituting the result in (\ref{3.asb}) gives
\begin{equation}\label{3.asb1}
\lim_{|\Lambda| \to \infty} {\Sigma_2(\Lambda) \over |\partial \Lambda|}
= {1 \over 8 \pi^{3/2}} \zeta ({3 \over 2})  = 0.0586436\cdots,
\end{equation}
where $\zeta(s)$ denotes the Riemann zeta function.
In the special case that $\Lambda$ is a disk of radius $r$, Prosen
\cite{10} has obtained a one dimensional integral representation of
$\Sigma_2(\Lambda)$, and from this obtained a numerical evaluation
consistent with (\ref{3.asb1}).

Another class of linear statistics of particular interest with regard
to the eigenvalues of random matrices (see e.g.~\cite{17} and
references therein) is when $a(\vec{r})$ varies on macroscopic length
scales relative to the spacing between particles. This can be achieved
by first supposing $a(\vec{r})$ varies on the length scale of the spacing
between zeros, and then making the replacement $a(\vec{r}) \mapsto
a(\vec{r}/\alpha)$ where $\alpha \gg 1$. Now, writing $S(\vec{r})$ 
in (\ref{3.for}) in terms
of its Fourier transform we have the formula
$$
{\rm Var}(A) = {1 \over (2 \pi)^2} \int_{{\bf R}^2} d\vec{r} \,
|\tilde{a}(k)|^2 \tilde{S}(\vec{k}).
$$
Making the replacement $a(\vec{r}) \mapsto
a(\vec{r}/\alpha)$ then gives
\begin{equation}\label{3.va}
{\rm Var}(A) = {\alpha^2 \over (2 \pi)^2} \int_{{\bf R}^2} d\vec{r} \,
|\tilde{a}(k)|^2 \tilde{S}(\vec{k}/\alpha).
\end{equation}
Since $\alpha \gg 1$ we see that the small $|\vec{k}|$ behaviour of
$\tilde{S}(\vec{k})$ determines ${\rm Var}(A)$. For the OCP
$\tilde{S}(\vec{k})$ exhibits the quadratic behaviour (\ref{3.lr1})
and so Var$(A)$ is formally independent of $\alpha$ in the macroscopic
limit \cite{16}. For the complex zeros, as already noted the fact
that the zeroth and second moments of $S(\vec{r})$ vanishes implies
that
\begin{equation}\label{3.va1}
\tilde{S}(\vec{k}) \mathop{\sim}\limits_{|\vec{k}| \to 0} c \vec{k}^4.
\end{equation}
Furthermore, from the definitions it is easy to see that $c$ is related
to the fourth moment of $\rho_{(2)}^T$ by
$$
c = {1 \over 64} \int_{{\bf R}^2}d\vec{r} \, \vec{r}^4 \rho_{(2)}^T(\vec{r},
\vec{0}).
$$
Using (\ref{3.h1}) this integral can be computed exactly, giving
$$
c = {1 \over 8 \pi} \zeta(3)
$$
(in fact all coefficients in the power series expansion of $\tilde{S}(\vec{k})$
have been calculated in \cite{4}).

Substituting (\ref{3.va1}) in (\ref{3.va}) we see that
\begin{equation}\label{3.va2}
{\rm Var}(A) 
 \mathop{\sim}\limits_{\alpha \to \infty}
\bigg ( {c \over (2 \pi)^2} \int_{{\bf R}^2} d\vec{k} \,
|\tilde{a}(k)|^2 |\vec{k}|^4 \bigg ) {1 \over \alpha^2},
\end{equation}
assuming the integral exists. (Note that the integral does not exist for
$a(\vec{r}) = \chi_{\Lambda_0}$, in keeping with the distinct behaviours
of (\ref{3.asb}) and (\ref{3.va2}).)

\subsection{Fluctuation formulas on the sphere}
The fluctuation formula (\ref{3.va2}) applies to the infinite system.
Also of interest is the situation of a fixed volume, and thus the
density being proportional to $N$ as the number of particles increases.
This is the natural setting for the zeros of the complex
random polynomial with variances specified by (\ref{1.5a})
projected onto the unit
sphere. We know from (\ref{3.sd1}) that the zeros are uniformly
distributed.
The question of interest is the behaviour of Var$(A)$ as a
function of $N$.

In fact this question, and its generalization to variance of
linear statistics for
zeros of multidimensional polynomials projected on to the
surface of higher dimensional spheres for which the zero density is
uniform, has been addressed recently by Shiffman and Zelditch
\cite{14}. They proved that for a statistic with $a(\theta,\phi)$
smooth on the unit sphere,
\begin{equation}\label{3.sz}
{\rm Var}(A) = O(1)
\end{equation}
as $N \to \infty$. This bound was also established in the higher
dimensional cases. The question now is if the $O(1)$ behaviour is
asymptotically exact. In light of the analysis for the infinite system
presented in the previous subsection, we would expect that asymptotically
exact $O(1)$ behaviour corresponds to $\tilde{S}(\vec{k}) \sim c_1 |\vec{k}|^2
$ in the infinite system. This can be seen from (\ref{3.va}) with
$\alpha^2$ identified with $N$. But $\tilde{S}(\vec{k})$ for the complex
zeros has the small $|\vec{k}|$ behaviour (\ref{3.va1}) which is
quartic in  $|\vec{k}|$. Identifying $\alpha^2$ with $N$ in (\ref{3.va2})
then suggests that
\begin{equation}\label{3.sz1}
{\rm Var}(A) \sim  {c' \over N}.
\end{equation}
We have not been able to prove this assertion in general, but it can be
demonstrated in a specific example, using a combination of analytic and
numerical analysis.

According to the formula (\ref{3.for}), our task is to compute
\begin{equation}\label{3.av}
{\rm Var}(A) = \int_{\Omega}dS_1 \int_{\Omega}dS_2 \, a(\hat{r}_1)
a(\hat{r}_2) \rho_{(2)}^T(\hat{r}_1,\hat{r}_2) +
\int_{\Omega} dS \, a^2(\hat{r}) \rho_{(1)}(\hat{r}),
\end{equation}
where $dS$ is the differential surface element for the unit sphere,
 $\rho_{(1)}(\hat{r})$ is given by (\ref{3.sd1})
and 
\begin{equation}\label{3.fud}
\rho_{(2)}^T(\hat{r}_1,\hat{r}_2) =  \rho_{(2)}^T(\hat{r}_1
\cdot\hat{r}_2)
\end{equation}
 is given by (\ref{3.sp}). To proceed further, we suppose
that in terms of the usual spherical coordinates $(\theta,\phi)$,
$$
 a(\hat{r}) = a(\cos\theta) = a(\hat{r}\cdot\hat{z})
$$
and thus depends only on the azimuthal angle. We want to use this
functional dependence, together with the functional dependence displayed
by (\ref{3.20'}) to simplify the double integral over the unit sphere in
(\ref{3.av}). 

First consider
\begin{equation}\label{3.i1}
I_1 := \int_{\Omega}dS_1 \, a(\hat{r}_1\cdot\hat{z})
\rho_{(2)}^T(\hat{r}_1
\cdot\hat{r}_2).
\end{equation}
In this integral we change variables $(\theta_1,\phi_1) \mapsto
(\theta,\phi)$ by rotating the sphere so that $\hat{r}_2 \mapsto
\hat{z}$, where now $\theta$ is measured as if the direction
$\hat{r}_2$ were the $z$-axis and $\phi = \phi_2$.
 Then, in the new coordinate frame
\begin{eqnarray*}
\hat{r}_2 & = & (0,0,1), \\
\hat{r}_1 & = & (\cos \phi \sin \theta, \sin \phi \sin \theta, \cos \theta), \\
\hat{z} & = & (-\cos \phi_2 \sin \theta_2, - \sin \phi_2 \sin \theta_2,
\cos \theta_2).
\end{eqnarray*}
Thus
\begin{eqnarray*}
\hat{r} \cdot \hat{z} & = & \cos \theta \cos \theta_2 - \cos \phi \cos
\phi_2 \sin \theta \sin \theta_2 - \sin \phi \sin \phi_2 \sin \theta
\sin \theta_2 \\
\hat{r}_1 \cdot \hat{r}_2 & = & \cos \theta
\end{eqnarray*}
and so
\begin{equation}\label{3.i2}
I_1 = \int_{\Omega}dS \, 
a(\cos \theta \cos \theta_2 - \cos \phi \cos \phi_2 \sin \theta \sin \theta_2
- \sin \phi \sin \phi_2 \sin \theta \sin \theta_2) \rho_{(2)}^T(\cos \theta).
\end{equation}

If we now further specialize and choose
\begin{equation}\label{3.sa}
a(\hat{r}_1\cdot\hat{z}) =  \hat{r}_1\cdot\hat{z}
\end{equation}
then, after recalling that 
\begin{equation}\label{3.ds}
dS = \sin \theta d\theta d\phi,
\end{equation}
 we see that
the integral over $\phi$ in (\ref{3.i2}) can be carried out, leaving
us with the single integral
$$
I_1 = 2 \pi \cos \theta_2 \int_0^\pi d\theta \, \sin \theta \cos \theta
\rho_{(2)}^T(\cos \theta).
$$
Substituting this simplification in (\ref{3.av}), together with
(\ref{3.sa}), (\ref{3.ds}) and (\ref{3.20'}) we thus obtain
\begin{eqnarray}\label{3.ff}
{\rm Var}\Big ( \sum_{j=1}^N \hat{r}_j \cdot\hat{z} \Big )
& = & 4 \pi^2 \int_0^\pi d\theta_2 \, \sin \theta_2
\cos^2\theta_2 \int_0^\pi d\theta \, \sin \theta
\cos \theta \rho_{(2)}^T(\cos \theta) \nonumber \\
&&+ {N \over 2} \int_0^\pi d\theta \, \sin \theta \cos^2 \theta \nonumber \\
& = & {8 \pi^2 \over 3} \int_0^\pi d\theta \, \sin \theta \cos \theta
\, \rho_{(2)}^T(\cos \theta) + {N \over 3}.
\end{eqnarray}
Although we have not been able to evaluate this integral analytically,
the exact formula (\ref{3.sp}) makes numerical integration
straightforward. The results are given in Table 1, and are consistent
with the expected behaviour (\ref{3.sz1}).

\begin{tabular}
{c|c c c|c} 
$N$ & Var$(A)$ & \hspace{2cm} & $N$ & Var$(A)$
\\[.2cm] 
10 & 0.1249 & & 80 & 0.0193 \\
20 & 0.0704 & & 90 & 0.0172 \\
30 & 0.0489 & & 100 & 0.0156 \\
40 & 0.0375 & & 110 & 0.0142 \\
50 & 0.0303 & & 120 & 0.0130 \\
60 & 0.0255 & & 130 & 0.0120 \\
70 & 0.0220 & & 140 & 0.0112
\end{tabular}

\vspace{.5cm}
\noindent 
{\bf Table 1. } Numerical values of (\ref{3.ff}), truncated
at the fourth decimal, exhibiting the expected
$O(1/N)$ decay of Var$(A)$.

\section{Correlations near the boundary of support}
As remarked in the Introduction, to leading order the density of the
complex zeros in the case of (\ref{1.5a}) is uniform in a disk of radius
$\sqrt{N}$ and zero outside that radius. The density of eigenvalues
implied by the p.d.f.~(\ref{1.6}) have the same feature, and therefore
so does the OCP at $\beta = 2$ with the Boltzmann factor (\ref{1.6})
taken to be
valid throughout the plane (the harmonic potential between background
and log-potential charge is only valid for the charge inside the
background; outside the potential must be $C+ N \log |\vec{r}|$ by
Newton's theorem). As the two-point function of the OCP has a
universal form at the boundary, it is of some interest to compute
the form of the one and two point functions for the complex zeros,
and compare their behaviour to that of the OCP.

\subsection{The density}
In the case that the variance of the coefficients
of the random polynomial (\ref{1.1}) is given by (\ref{1.5a}),
we have noted that
the density of zeros is given by  (\ref{3.den1}). Here we want to analyze
$\rho_{(1)}(\vec{r})$ with $\vec{r}$ measured from the boundary of
support of the density $|\vec{r}|=\sqrt{N}$. This can be achieved by
writing
\begin{equation}\label{4.2}
z = x + i (y - \sqrt{N}),
\end{equation}
which sets the origin as the southern boundary of the support. 
According to (\ref{3.den1}) the asymptotic expansion of
$e(|z|^2;N)$ with $z$ a function of $N$ as specified by (\ref{4.2})
is required. For this purpose we note from the definition (\ref{3.den2})
the easily verified formula 
$$
e(z;N) = e^z {\Gamma (1 + N;z) \over \Gamma (1 + N)}, \quad
\Gamma(N;a) := \int_a^\infty e^{-t} t^{N-1} \, dt,
$$
and the asymptotic expansion \cite{23.1}
\begin{equation}\label{4.3}
{\Gamma(1+N;N + \sqrt{2N} u) \over \Gamma (1 + N)} =
{1 \over 2} \Big ( 1 - {\rm erf} (u) + O({1 \over \sqrt{N}}) \Big ).
\end{equation}
Choosing $u = (z - N)/\sqrt{2N}$ we see that
\begin{equation}\label{4.4}
e(z;N) = {1 \over 2} e^z \Big (1 - {\rm erf} \Big ( {z - N \over \sqrt{2N}}
\Big ) + O({1 \over \sqrt{N}}) \Big ),
\end{equation}
valid for $(z - N)/\sqrt{2N}$ bounded, which in turn implies
\begin{equation}\label{4.5}
e(|x + i(y - \sqrt{N})|^2;N) \sim {1 \over 2} e^{|z|^2}
\Big ( 1 + {\rm erf}(\sqrt{2} y) \Big ),
\end{equation}
where $z$ is given by (\ref{4.2}). Hence, with the limiting form of
$\rho_{(1)}(\vec{r})$ denoted by $\rho_{(1)}(y)$, we see by
substituting (\ref{4.5}) in (\ref{3.den1}) that
\begin{equation}\label{4.6}
\pi \rho_{(1)}(y) = 1 + {1 \over 4} {\partial^2 \over \partial y^2}
\log \Big ( 1 + {\rm erf} (\sqrt{2} y) \Big ).
\end{equation}

A plot of $\pi \rho_{(1)}(y)$, which is most expediently carried out
by first explicitly calculating the second derivative to give
\begin{equation}\label{4.7}
\pi \rho_{(1)}(y) = 1 - {\sqrt{2 \over \pi} 2y e^{-2y^2}
(1 + {\rm erf}(\sqrt{2}y) ) + {2 \over \pi} e^{-4y^2} \over
(1 + {\rm erf} (\sqrt{2} y ))^2},
\end{equation}
indicates rapid approaching 
of the bulk limit $\pi\rho_{(1)}(y) = 1$ as $y \to \infty$, but slow
decay outside the boundary of support as $y \to -\infty$. This
observation is easily confirmed analytically from (\ref{4.6}). Use of 
the expansion
$$
1 + {\rm erf} (\sqrt{2} y) \sim \left \{
\begin{array}{ll}
1 + {e^{-2y^2} \over \sqrt{2 \pi} y}, & y \to \infty \\[.2cm]
 {e^{-2y^2} \over \sqrt{2 \pi} |y|}, & y \to - \infty \end{array}
\right.
$$
shows 
\begin{equation}\label{4.8}
\pi \rho_{(1)}(y)
\sim \left \{
\begin{array}{ll}
1 + 4 {1  \over \sqrt{2 \pi}} y e^{-2y^2}, & y \to \infty \\[.2cm]
{1 \over 4 y^2}, & y \to - \infty \end{array}
\right.
\end{equation}

\begin{figure}
\epsfxsize=10cm
\centerline{\epsfbox{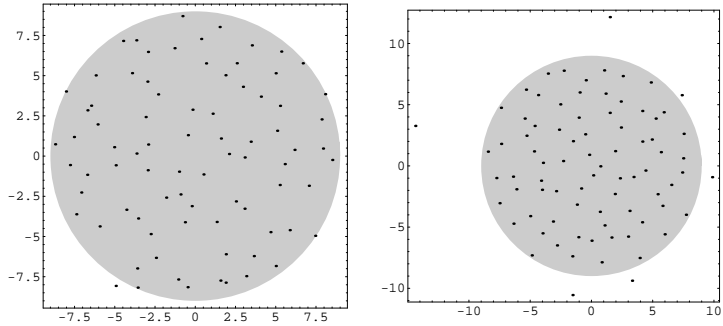}}
\caption{\label{f4.2} A typical realization of the eigenvalues of a
$81 \times 81$
complex Gaussian random matrix (leftmost plot)
and the zeros of a complex Gaussian random polynomial 
of degree 81 with variances given by (\ref{1.5a}). The shaded region
represents the disk $|z| <9$ which is the leading order support
of the density in both cases. Outside the disk the density has a
$1/r^2$ tail in the case of the zeros, whereas it falls off as
a Gaussian for the eigenvalues, in keeping with the realizations
in the Figure.}
\end{figure}

Consider now the density for the OCP with Boltzmann factor proportional
to (\ref{1.6}). It is given by the exact formula \cite{21}
\begin{equation}\label{4.cr1}
\pi \rho_{(1)}(\vec{r}) = e^{-|\vec{r}|^2} e(|z|^2;N),
\end{equation}
so we see from (\ref{4.5}) that in the limit $N \to \infty$ the density
as measured from the southern most point on the boundary of support is
given by
\begin{equation}\label{4.9}
\pi \rho_{(1)}(y) = h(y), \qquad
h(y) := {1 \over 2} \Big ( 1 + {\rm erf}(\sqrt{2}y) \Big ).
\end{equation}
In contrast to the edge density (\ref{4.7}) which exhibits algebraic
decay as seen in (\ref{4.8}) for large distances outside the boundary of
support, here the asymptotic values are reached with only Gaussian
corrections in both directions,
$$
\pi \rho_{(1)}(y)
\sim \left \{
\begin{array}{ll}
1 -  {e^{-2y^2}  \over 2 \sqrt{2 \pi} |y|}, & y \to \infty \\[.2cm]
{e^{-2y^2} \over  2 \sqrt{2 \pi} |y|}, & y \to - \infty \end{array}
\right.
$$
The sharpness of the boundary of support for the 
density associated with (\ref{1.6}), as generated from the eigenvalues
of complex random matrices, relative to that for the zeros of
complex random matrices can be illustrated by plotting the zeros for
a particular realization in each case. This is done in Figure \ref{f4.2}.

\vspace{.5cm}

Another point of interest is that analogous behaviour to (\ref{4.8})
is found for the OCP at $\beta = 2$ with the correct Coulomb potential 
$C+N\log r$ between 
particles and background outside the disk. The exact density as measured
from the southern most boundary of the uniform background is given by
\cite{24}
\begin{equation}\label{4.ja1}
\rho_{(1)}(y) = f(y) {2 \over \sqrt{\pi}} \int_0^\infty
{\exp(2^{3/2}tx - t^2) \over 1 + {\rm erf}(t) + e^{-t^2}/t \sqrt{\pi}}
\end{equation}
where $f(y) = e^{-2y^2}$, $y > 0$ and $f(y)=1$, $y<0$. This gives for the
asymptotic behaviour \cite{24}
\begin{equation}\label{4.8'}
\pi \rho_{(1)}(y)
\sim \left \{
\begin{array}{ll}
1 -  {e^{- 2 y^2}  \over 2 \sqrt{2 \pi} |y|^3}, & y \to \infty \\[.2cm]
{1 \over  4 y^2}, & y \to - \infty \end{array}
\right.
\end{equation}

An analytic feature of the surface density for the OCP in general is that
the total charge is zero. This means
\begin{equation}\label{4.ss}
\int_{-\infty}^0 \pi \rho_{{\rm OCP}(1)}(y) \, dy +
\int_0^\infty \pi ( \rho_{{\rm OCP}(1)}(y) - 1) \, dy = 0,
\end{equation}
which indeed can be verified for the OCP profiles (\ref{4.cr1}) and
(\ref{4.ja1}). A straightforward calculation using (\ref{4.6}) also
shows that (\ref{4.ss}) holds true for the boundary density of the
complex zeros.

\subsection{The two-point distribution}
Let $\rho_{(2)}(y_1,y_2;x_1-x_2)$ denote the $N \to \infty$ limit of
$\rho_{(2)}(\vec{r}_1,\vec{r}_2)$ as specified by (\ref{3.den4'})
with
\begin{equation}\label{4.10}
z_1 = x_1 + i (y_1 - \sqrt{N}), \qquad z_2 = x_2 + i (y_2 - \sqrt{N}).
\end{equation}
The basic strategy is to explicitly compute the limiting value of the
matrix elements of
\begin{equation}\label{4.U}
U := dC - B^\dagger A^D B
\end{equation}
and the determinant $d$ in (\ref{3.den4'}). From (\ref{3.den6}) and 
(\ref{3.den6'}) we have
$$
d = a_{11} a_{22} - a_{12}a_{21} = 
e(|z_1|^2;N) e(|z_2|^2;N) - |e(z_1\bar{z}_2;N) |^2.
$$
With $z_1$ and $z_2$ given by (\ref{4.10}) we see that the asymptotics
of the first term follows from (\ref{4.5}), while (\ref{4.4}) shows
\begin{equation}\label{4.11}
e(z_1\bar{z}_2;N) \sim {1 \over 2} e^{z_1 \bar{z}_2}
\Big \{ 1 + {\rm erf} \Big ( {1 \over \sqrt{2}} (y_1 + y_2 - i (x_1 - x_2)
\Big ) \Big \}.
\end{equation}
Thus, with $h(y)$ defined as in (\ref{4.9}) we have
\begin{eqnarray}\label{4.12}
d & \sim & e^{2N - 2 \sqrt{N} (y_1 + y_2)} e^{(x_1^2 + x_2^2 + y_1^2 + y_2^2)}
\bigg ( h(y_1) h(y_2) \nonumber \\
&& - e^{-(x_1 - x_2)^2 - (y_1 - y_2)^2} \Big |
h\Big ( {1 \over 2} (y_1 + y_2 - i (x_1 - x_2)) \Big ) \Big |^2 \bigg ).
\end{eqnarray}

The analysis of the leading order behaviour of the matrix elements of
(\ref{4.U}) is more involved. Each element consists of the sum of
many individual elements, and we find that a naive leading order
expansion of each term individually gives zero, as all such contributions
exactly cancel. It is therefore necessary to expand individual terms to
higher order. This can conveniently be achieved by writing the terms 
$e(z;N+p)$ which occur therein with $p \ne 0$ in terms of $e(z;N)$.
For example, from the definition (\ref{3.den2}) we have
\begin{equation}\label{4.13}
e(z;N-1) = e(z;N) - {z^N \over N!}, \quad
e(z;N-2) = e(z;N) - {z^N \over N!} - {z^{N-1} \over (N-1)!}
\end{equation}
which we use to substitute for $e(z;N-1)$ and $e(z;N-2)$.

Let us give the details of the implementation of this procedure in the case
of $u_{11}$. First, according to (\ref{4.U}), we have
$$
u_{11} = d c_{11} + \bar{b}_{11} a_{12} b_{21} +
\bar{b}_{21} a_{21} b_{11} - \bar{b}_{11} a_{22} b_{11} -
\bar{b}_{21} a_{11} b_{21}.
$$
Substituting in the explicit form of the matrix elements according to
(\ref{3.den6'}) then gives
\begin{eqnarray*}
u_{11} & = & \Big (|z_1|^2 e(|z_1|^2;N-2) + e(|z_1|^2;N-1) \Big )
\Big ( e(|z_1|^2;N) e(|z_2|^2;N) - |e(z_1 \bar{z}_2;N) |^2 \Big ) \\
&& - |z_1|^2 | e(|z_1|^2;N-1)|^2 e(|z_2|^2;N)
- |z_2|^2|e(z_1 \bar{z}_2;N-1)|^2 e(|z_1|^2;N) \\
&&+2 {\rm Re} \Big (
z_1 \bar{z}_2 e(z_1 \bar{z}_2;N-1) e(\bar{z}_1 z_2;N) e(|z_1|^2;N-1) \Big ).
\end{eqnarray*}

With the susbstitution (\ref{4.13}) this can be rewritten to read
$$
u_{11} = (a) + (b) + (c) + (d) + (e) + (f) + (g1) + (g2) + (g3),
$$
where
\begin{eqnarray*}
(a) & := & | e(|z_1|^2;N)|^2 e(|z_2|^2;N) \\
(b) & := & |e(z_1\bar{z}_2;N)|^2  e(|z_1|^2;N)\Big ( - 1 - |z_1|^2
- |z_2|^2 + z_1 \bar{z}_2 + z_2 \bar{z}_1 \Big ) \\
(c) & := &  e(|z_1|^2;N)  e(|z_2|^2;N)
\Big ( - {|z_1|^{2N} \over N!} + {|z_1|^{2N} \over N!}(|z_1|^2 - N) \Big ) \\
(d) & := & |e(z_1\bar{z}_2;N)|^2 {|z_1|^{2N} \over N!}
\Big (|z_1|^2 + N + 1 - z_1 \bar{z}_2 - \bar{z}_1 z_2 \Big ) \\
(e) & := & e(|z_1|^2;N) e(z_1\bar{z}_2;N) {(\bar{z}_1 z_2)^N \over N!}
\Big (|z_2|^2 - \bar{z}_1 z_2 \Big ) \\
(f) & := & e(|z_1|^2;N)e(\bar{z}_1 z_2;N){({z}_1 \bar{z}_2)^N \over N!}
\Big (|z_2|^2 - {z}_1 \bar{z}_2 \Big ) \\
(g1) & := & - {|z_1|^{4N + 2} \over (N!)^2} e(|z_2|^2;N) \\
(g2) & := & - {|z_1|^{2N}|z_2|^{2N+2} \over (N!)^2} e(|z_1|^2;N) \\
(g3) & := & 2 {|z_1|^{2N} \over N!} {\rm Re} \Big (
{(z_1 \bar{z}_2)^{N+1} \over N!} e(\bar{z}_1{z}_2;N) \Big ).
\end{eqnarray*}
It now remains to calculate the large $N$ asymptotic behaviour of each of
these terms with $z_1$ and $z_2$ given by (\ref{4.10}). Using (\ref{4.5}),
(\ref{4.11}) and the expansion 
\begin{eqnarray*}
{(z_i \bar{z}_j)^N \over N!} & \sim & {1 \over \sqrt{2 \pi N}}
\exp \Big ( N - \sqrt{N} (y_i + y_j + i (x_j - x_i) \Big ) \\
&& + {1 \over 2}(x_i^2 + x_j^2 - y_i^2 - y_j^2) + i (x_i y_i - x_j y_j)
\Big )
\end{eqnarray*}
which follows from the expansion $(1+u)^N \sim e^{N(u - u^2/2 + \cdots)}$
and Stirling's formula, we obtain
\begin{eqnarray*}
(a) & \sim & e^{3N - 4 \sqrt{N} y_1 - 2\sqrt{N} y_2}
\exp \Big ( 2(x_1^2 + y_1^2) + x_2^2 + y_2^2 \Big ) h^2(y_1) h(y_2) \\
(b) & \sim & e^{3N - 4 \sqrt{N} y_1 - 2\sqrt{N} y_2}
\Big ( - (x_1 - x_2)^2 - (y_1 - y_2)^2 - 1 \Big ) \exp \Big ( x_1^2 + y_1^2 +
2x_1 x_2 + 2 y_1 y_2 \Big ) \\
&& \times h(y_1) \Big | h\Big ( {1 \over 2}(y_1 + y_2 + i (x_1 - x_2))
\Big ) \Big |^2 \\
(c) & \sim & - e^{3N - 4 \sqrt{N} y_1 - 2\sqrt{N} y_2}
\sqrt{2 \over \pi} y_1 \exp (2x_1^2 + x_2^2 + y_2^2) h(y_1) h(y_2) \\
(d) & \sim &  e^{3N - 4 \sqrt{N} y_1 - 2\sqrt{N} y_2}
\sqrt{2 \over \pi} y_2 \exp( x_1^2 - y_1^2 + 2(x_1 x_2 + y_1 y_2)) \\
&& \times \Big | h\Big ( {1 \over 2}(y_1 + y_2 + i (x_1 - x_2))
\Big ) \Big |^2 \\
(e) + (f) & \sim & e^{3N - 4 \sqrt{N} y_1 - 2\sqrt{N} y_2}
\exp \Big ( {1 \over 2}(3x_1^2 + y_1^2 + x_2^2 - y_2^2) + x_1 x_2 + y_1 y_2 
\Big ) h(y_1) \\
&& \times 2 {\rm Re}\Big \{ {1 \over \sqrt{2 \pi}}(y_1 - y_2 + i(x_1 - x_2))
\exp\Big ( i (x_2 y_2 - x_1 y_1 + x_2 y_1 - x_1 y_2) \Big ) \nonumber \\
&& \times  h\Big ( {1 \over 2}(y_1 + y_2 - i (x_1 - x_2))
\Big ) \Big \} \\
(g1) & \sim & -  e^{3N - 4 \sqrt{N} y_1 - 2\sqrt{N} y_2}
{1 \over 2 \pi} \exp(2x_1^2 - 2y_1^2 + x_2^2 + y_2^2) h(y_2) \\
(g2) & \sim & -  e^{3N - 4 \sqrt{N} y_1 - 2\sqrt{N} y_2}
{1 \over 2 \pi} \exp(2x_1^2 + x_2^2 - y_2^2) h(y_1) \\
(g3) & \sim &   e^{3N - 4 \sqrt{N} y_1 - 2\sqrt{N} y_2}
{1 \over2 \pi}\exp \Big ( {1 \over 2}(3x_1^2 - 3 y_1^2 + x_2^2 - y_2^2) + x_1 x_2 + y_1 y_2 
\Big ) \\
&& \times 2 {\rm Re}\Big \{ 
\exp\Big ( i (x_1 y_1 - x_2 y_2 + x_1 y_2 - x_2 y_1) \Big )
 h\Big ( {1 \over 2}(y_1 + y_2 + i (x_1 - x_2))
\Big ) \Big \}.
\end{eqnarray*}

Notice that all of the above asymptotic forms contain an $N$ dependent
factor $e^{3N - 4\sqrt{N}y_1 - 2\sqrt{N} y_2}$. Similarly decomposing
$u_{22}$ shows that the asymptotic form of each term contains the $N$
dependent factor $e^{3N - 4\sqrt{N}y_2 - 2\sqrt{N} y_1}$. Now, according to
(\ref{3.den6}) and (\ref{3.den4'})
$$
\pi^2 \rho_{(2)}(\vec{r}_1,\vec{r}_2) = {u_{11} u_{22} \over d^3} +
{u_{12} u_{21} \over d^3}
$$
so we are required to form the product $u_{11} u_{22}$ and to divide by
$d^3$. Doing this, we see that the $N$ dependent factors cancel out as
expected. Similarly we find that $N$ dependent factors in $u_{12} u_{21}/
d^3$ cancel. Collecting together the remaining terms, which are
calculated for $u_{22}$, $u_{12}$ and $u_{21}$ in the same manner as
detailed above for $u_{11}$, we obtain for the limiting value of the
edge two-point distribution
\begin{equation}\label{4.pv}
\pi^2 \rho_{(2)}(y_1,y_2;x_1 - x_2) = {v_{11} v_{22} \over d_1^3} +
{v_{12} v_{21} \over d_2^3}
\end{equation}
where
\begin{eqnarray}
d_1 & := & h(y_1) h(y_2) - 
e^{-(x_1 - x_2)^2 - (y_1 - y_2)^2} \Big |
 h\Big ( {1 \over 2}(y_1 + y_2 + i (x_1 - x_2)) \Big )\Big |^2 \label{4.a1} \\
d_2 & := & e^{(x_1 - x_2)^2 + (y_1 - y_2)^2} d_1 \label{4.A}\\
v_{11} & = & h^2(y_1) h(y_2) - \Big (
(x_1 - x_2)^2 + (y_1 - y_2)^2 + 1 \Big ) e^{-(x_1 - x_2)^2 - (y_1 - y_2)^2}
h(y_1) \nonumber \\
&&\times  \Big |  h\Big ( {1 \over 2}(y_1 + y_2 + i (x_1 - x_2)) \Big )\Big |^2
- \sqrt{2 \over \pi} y_1 e^{-2y_1^2} h(y_1) h(y_2) \nonumber \\
&& + \sqrt{2 \over \pi} y_2 e^{-2y_1^2}
e^{-(x_1 - x_2)^2 - (y_1 - y_2)^2} \Big |
 h\Big ( {1 \over 2}(y_1 + y_2 + i (x_1 - x_2)) \Big )\Big |^2 \nonumber \\
&& +2 {\rm Re} \Big \{ {1 \over \sqrt{2 \pi}}(y_1 - y_2 + i (x_1 - x_2))
e^{-(y_1^2 + y_2^2)} 
e^{-(x_1 - x_2)^2/2 - (y_1 - y_2)^2/2} \nonumber \\
&&\times e^{i(x_2 - x_1) (y_1 + y_2)} h(y_1) 
 h\Big ( {1 \over 2}(y_1 + y_2 - i (x_1 - x_2)) \Big \} \nonumber \\
&& - {1 \over 2 \pi} e^{-4y_1^2} h(y_2) - {1 \over 2 \pi}
e^{-2 (y_1^2 + y_2^2)} h(y_1) + 2 {\rm Re} \Big \{ {1 \over 2 \pi}\nonumber \\
&& 
\times
e^{-3y_1^2 - y_2^2} e^{-(x_2 - x_1)^2/2 - (y_2 - y_1)^2/2} e^{i(x_1 - x_2)
(y_1 + y_2)} 
h\Big ( {1 \over 2}(y_1 + y_2 + i (x_1 - x_2)) \Big )\Big ) \Big \}
\label{4.a2} \\
v_{22} & = & v_{11} \Big |_{x_1 \leftrightarrow x_2 \atop y_1
\leftrightarrow y_2} \label{4.a}\\
v_{12} & = & - \Big (
(x_1 - x_2)^2 + (y_1 - y_2)^2 - 1 \Big ) e^{(x_1 - x_2)^2 + (y_1 - y_2)^2}
h(y_1) h(y_2 )  h\Big ( {1 \over 2}(y_1 + y_2 - i (x_1 - x_2)) \Big )
\nonumber \\
&& - 
 h\Big ( {1 \over 2}(y_1 + y_2 - i (x_1 - x_2)) \Big )
\Big |
 h\Big ( {1 \over 2}(y_1 + y_2 + i (x_1 - x_2)) \Big )\Big |^2
\nonumber \\
&& -  {1 \over \sqrt{2 \pi}}(y_1 + y_2 + i (x_1 - x_2))
e^{-(y_1^2 + y_2^2)} e^{3(x_2 - x_1)^2/2 + 3(y_2 - y_1)^2/2}
e^{i(x_1 - x_2)(y_1 + y_2)} h(y_1) h(y_2) \nonumber \\
&& + {1 \over \sqrt{2 \pi}}(y_1 + y_2 - i (x_1 - x_2))e^{-(y_1^2 + y_2^2)}
e^{(x_1 - x_2)^2/2 + (y_1 - y_2)^2/2} 
\nonumber \\
&&\times e^{i(x_1 - x_2)(y_1 + y_2)}
\Big |
 h\Big ( {1 \over 2}(y_1 + y_2 + i (x_1 - x_2)) \Big )\Big |^2 \nonumber
\\
&& + {1 \over \sqrt{2 \pi}} \Big ( y_1 - y_2 + i(x_1 - x_2) \Big )
e^{-2y_2^2} e^{(x_2 - x_1)^2 + (y_2 - y_1)^2}  h(y_1)
 h\Big ( {1 \over 2}(y_1 + y_2 - i (x_1 - x_2)) \Big ) \nonumber 
\\&& + {1 \over \sqrt{2 \pi}} \Big ( y_2 - y_1 + i(x_1 - x_2) \Big )
e^{-2y_2^1} e^{(x_2 - x_1)^2 + (y_2 - y_1)^2}  h(y_2)
 h\Big ( {1 \over 2}(y_1 + y_2 - i (x_1 - x_2)) \Big ) \nonumber 
\\&& + {1 \over 2 \pi} e^{-2(y_1^2 + y_2^2)} e^{(x_2 - x_1)^2 + (y_2 - y_1)^2}
e^{2 i (x_1 - x_2) (y_1 + y_2)}  
 h\Big ( {1 \over 2}(y_1 + y_2 + i (x_1 - x_2)) \Big )\nonumber \\
&&+ {1 \over 2 \pi} e^{-2(y_1^2 + y_2^2)} e^{(x_2 - x_1)^2 + (y_2 - y_1)^2}
 h\Big ( {1 \over 2}(y_1 + y_2 - i (x_1 - x_2)) \Big ) \nonumber \\
&&- {1 \over 2 \pi} e^{-3y_1^2 - y_2^2} e^{3(x_2 - x_1)^2/2 +
3(y_2 - y_1)^2/2} e^{i(x_1 - x_2) (y_1 + y_1)} h(y_2) \nonumber \\
&&- {1 \over 2 \pi} e^{- y_1^2 - 3 y_2^2} e^{3(x_2 - x_1)^2/2 +
3(y_2 - y_1)^2/2} e^{i(x_1 - x_2) (y_1 + y_1)} h(y_1)
\label{4.a3} \\
v_{21} & = & \bar{v}_{12}. \label{4.b}
\end{eqnarray}

The above explicit forms show that indeed $\pi^2 \rho_{(2)}(y_1,y_2;x_1 - 
x_2)$ as specified by (\ref{4.pv}) is a function of $x_1 - x_2$ as
claimed in the notation. Physically, this corresponds to translation
invariance in the direction of the boundary. Another check on
(\ref{4.pv}) is that in the limit $y_1,y_2 \to \infty$,
$y_1 - y_2$ fixed, the bulk result (\ref{3.bt}) is reclaimed. In this
limit, since erf$(u) \to 1$ as $u \to \infty$ we have $h(u) \to 1$ as
$u \to \infty$. This shows that
\begin{equation}
\begin{array}{ll}
v_{11} \sim v_{22} \sim 1 - (r^2 + 1) e^{-r^2}, &
v_{12} \sim v_{21} \sim -(r^2 - 1) e^{r^2} - 1 \\
d_1 \sim 1 - e^{-r^2}, & d_2 \sim e^{r^2} - 1
\end{array}
\end{equation}
where $r^2 := (x_1 - x_2)^2 + (y_1 - y_2)^2$, and hence
$$
\pi^2 \rho_{(2)}(y_1,y_2;x_1 - 
x_2) \sim {1 \over (1 - e^{-r^2})^3}
\Big ( e^{-2 r^2}(e^{r^2} - 1 - r^2)^2 + e^{-r^2}(e^{-r^2} - 1 + r^2)^2 \Big )
$$
which is (\ref{3.bt}) in the form given by Prosen \cite[eq.~(27)]{10}.

It is of interest to compare the structure of (\ref{4.pv}) with the
limiting value of the edge two-point distribution function for the
OCP with Boltzmann factor (\ref{1.6}). In the finite system the
exact two-point distribution is given by \cite{21}
\begin{equation}\label{4.cr2}
\pi^2 \rho_{(2)}(\vec{r}_1,\vec{r}_2) = \pi^2
\rho_{(1)}(\vec{r}_1) \rho_{(1)}(\vec{r}_2) - \pi^2
e^{-|\vec{r}_1|^2 - |\vec{r}_2|^2} |e(z_1 \bar{z}_2;N) |^2
\end{equation}
where $\pi \rho_{(1)}(\vec{r})$ is specified by (\ref{4.cr1}). Use of
(\ref{4.11}) then gives that in the limit $N \to \infty$ with the
coordinates (\ref{4.10}) we have
\begin{equation}\label{4.cr3}
\pi^2 \rho_{(2)}(y_1,y_2;x_1 - 
x_2)  = \pi^2 \rho_{(1)}(y_1) \rho_{(1)}(y_2) - \pi^2
e^{-(x_1 - x_2)^2 - (y_1 - y_2)^2} \Big |
\rho_{(1)} \Big ({1 \over 2}(y_1 + y_2 - i (x_1 - x_2)) \Big ) \Big |^2
\end{equation}
where $\pi \rho_{(1)}(y)$ is specified by (\ref{4.9}). In (\ref{4.cr3})
the general structure of the two-point distribution, as a product of
the one-body densities plus a correction term (the truncated two-point
distribution, or two-point correlation function $\rho_{(2)}^T$) which
decays as the points $(x_1,y_1)$ and $(x_2,y_2)$ are separated, is displayed.
Thus we have
\begin{equation}\label{4.cr4}
\rho_{(2)}^T(y_1,y_2;x_1 - 
x_2) = - e^{-(x_1 - x_2)^2 - (y_1 - y_2)^2} \Big |
\rho_{(1)} \Big ({1 \over 2}(y_1 + y_2 - i (x_1 - x_2)) \Big ) \Big |^2.
\end{equation}
This is distinct from the exact expression (\ref{4.pv}) in which
such a decomposition is not exhibited.

\subsection{Asymptotics of the two-point correlation}
In this subsection the asymptotic behaviour of the two-point correlation
parallel to the boundary of support of the density will be
investigated. In the case of the two-point correlation (\ref{4.cr3})
for the OCP at $\beta = 2$, this behaviour must conform to a universal
form obeyed by the asymptotics  of the charge-charge correlation
parallel to a plane boundary for general two-dimensional Coulomb systems
in their conducting phase. This asymptotic behaviour in the case of the
OCP specifies that \cite{25}
\begin{equation}\label{4.uf}
\rho_{{\rm OCP} \,(2)}^T(y_1,y_2;x_1 - 
x_2) \mathop{\sim}\limits_{|x_1 - x_2| \to \infty} - {f(y_1,y_2) \over
2 \beta\pi^2 (x_1 - x_2)^2}, \qquad \int_{-\infty}^\infty dy_1
\int_{-\infty}^\infty dy_2 \, f(y_1,y_2) = 1.
\end{equation}
To test the prediction (\ref{4.uf}) on the exact result (\ref{4.cr3}), the
asymptotic expansion of the error function 
erf$(Y + i X)$ for large imaginary argument
is required. Now, by choosing the path of integration from 0 to
$Y + i X$ in the definition of erf$(Y + i X)$ to first go along the
imaginary axis to $iX$ then parallel to the real axis to $Y + iX$, we
see that it is possible to write
$$
{\rm erf}(Y + i X) = i {2 \over \sqrt{\pi}} \int_0^X e^{s^2} \, ds +
{2 \over \sqrt{\pi}} e^{X^2} \int_0^Y e^{-2i X s - s^2} ds.
$$
Using integration by parts we can then deduce that
\begin{equation}\label{4.ee}
{\rm erf}(Y + i X) \mathop{\sim}\limits_{X \to \infty} {1 \over \sqrt{\pi}}
e^{X^2 - Y^2 - 2i XY} \Big ( {i \over X} - {Y \over X^2}
+ {i \over 2 X^3} (1 - 2Y^2) - {Y(3 - 2Y^2) \over 2 X^4}
+ O({1 \over X^5}) \Big ).
\end{equation}
This in turn, when used in (\ref{4.cr3}), implies
\begin{equation}\label{4.rf}
\rho_{{\rm OCP} \,(2)}^T(y_1,y_2;x_1 - 
x_2) \mathop{\sim}\limits_{|x_1 - x_2| \to \infty} -
{f(y_1) f(y_2) \over 4 \pi^2(x_1 - x_2)^2}, \qquad
f(y) := {d \over dy} \pi \rho_{(1)}(y)
\end{equation}
which indeed obeys (\ref{4.uf}).

Let us now consider the asymptotic expansion of (\ref{4.pv}) for
$|x_1 - x_2| \to \infty$. Since this expression represents the full
two-point distribution rather than its truncated counterpart we expect
the leading behaviour to be given by $\pi^2 \rho_{(1)}(y_1) \rho_{(1)}(y_2)$,
where each $\rho_{(1)}(y)$ is specified by (\ref{4.7}). The next term
will then give the leading order behaviour of $\rho_{(2)}^T$. Now
according to (\ref{4.pv}) and (\ref{4.A}), (\ref{4.a}), (\ref{4.b})
the asymptotic behaviour follows from the behaviour of
the quantities $d_1$, 
$v_{11}$ and $v_{12}$. Inspection of (\ref{4.a1}), (\ref{4.a2}) and
(\ref{4.a3}) shows that the latter asymptotic behaviour
follows from that of $h({1 \over 2}(y_1 + y_2 - i(x_1 - x_2)))$, its
complex conjugate and its modulus squared. Now, from the definition
(\ref{4.9}) and the expansion  (\ref{4.ee}) we have that
\begin{eqnarray*}
h\Big ({1 \over 2}(y_1 + y_2 - i(x_1 - x_2)\Big ) & \sim &
{1 \over \sqrt{2 \pi}} e^{{1 \over 2}(x_1 - x_2)^2 - {1 \over 2}
(y_1 + y_2)^2 - i (x_2 - x_1)(y_1 + y_2)} \nonumber \\
&& \times \Big ( {i \over x_2 - x_1} - {y_1 + y_2 \over (x_2 - x_1)^2}
+ {i \over (x_2 - x_1)^3}(1 - (y_1 + y_2)^2) + \cdots \Big )
\\
h\Big ({1 \over 2}(y_1 + y_2 + i(x_1 - x_2)\Big ) & \sim &
\overline{h\Big ({1 \over 2}(y_1 + y_2 - i(x_1 - x_2)\Big )}
\\
\Big | h\Big ({1 \over 2}(y_1 + y_2 - i(x_1 - x_2)\Big ) 
\Big |^2 & \sim &
{1 \over 2 \pi} e^{(x_1 - x_2)^2 - (y_1 + y_2)^2} \Big ( {1 \over (x_1 - x_2)^2}
\\ && + {1 \over (x_1 - x_2)^4}(2 - (y_1 + y_2)^2) + \cdots \Big )
\end{eqnarray*}
After some calculation we then find
\begin{eqnarray}
d_1 & \sim & h(y_1) h(y_2) \Big ( 1 - {e^{-2(y_1^2 + y_2^2)} \over 2 \pi
h(y_1) h(y_2)} {1 \over (x_1 - x_2)^2} \Big ) \label{4.1i} \\
v_{11} & \sim & h(y_2) \Big ( h^2(y_1) - \sqrt{2 \over \pi} y_1
e^{-2y_1^2} h(y_1) - {1 \over 2 \pi} e^{-4y_1^2} \Big ) \nonumber \\
&& + {e^{-2(y_1^2 + y_2^2)} \over (x_1 - x_2)^2}
\Big ( - {1 \over 2 \pi} h(y_1) (1 + 4y_1^2) - {1 \over \sqrt{2 \pi^3}}
y_1 e^{-2y_1^2} \Big ) \label{4.2i} \\
v_{12}  & \sim & e^{{3 \over 2} (x_1 - x_2)^2}
e^{-{3 \over 2} (y_1 + y_2)^2} e^{-i(x_2 - x_1)(y_1 + y_2)}
\nonumber \\
&&\bigg \{ {i \over x_2 - x_1} \Big ( {1 \over \sqrt{2 \pi}} 4y_1 y_2
h(y_1) h(y_2) e^{2 (y_1^2 + y_2^2)} + {1 \over \pi} y_1 h(y_1) e^{2y_1^2}
+ {1 \over \pi} y_2 h(y_2) e^{2y_2^2} + {1 \over \sqrt{8 \pi^3}} \Big )
\nonumber \\
&& + {1 \over (x_2 - x_1)^2}\Big ( {2 \over \sqrt{2 \pi}} (y_1 + y_2)
(1 - 2y_1 y_2) h(y_1) h(y_2) e^{2 (y_1^2 + y_2^2)}
+ {1 \over 2 \pi} (1 - 2y_2 (y_1 + y_2))  \nonumber \\
&& \times h(y_2) e^{2 y_2^2} 
+ {1 \over 2 \pi} (1 - 2y_1 (y_1 + y_2)) h(y_1) e^{2 y_1^2} -
{1 \over \sqrt{8 \pi^3}}(y_1 + y_2) \Big ) \bigg \}
\label{4.3i}
\end{eqnarray}

Substituting the results (\ref{4.1i}), (\ref{4.2i}) and (\ref{4.3i}) in
(\ref{4.pv}), recalling the relations (\ref{4.A}), (\ref{4.a}) and
(\ref{4.b}), and performing some further calculation we find the sought
expansion,
\begin{equation}
\pi^2 \rho_{(2)}(y_1,y_2;x_1 - x_2) \sim {g(y_1) \over h^2(y_1)}
{g(y_2) \over h^2(y_2)} + {Q(y_1,y_2) \over (x_1 - x_2)^2}
\end{equation}
where
$$
g(y) := h^2(y) - \sqrt{2 \over \pi} y e^{-2y^2} h(y) - {1 \over 2 \pi}
e^{-4y^2}
$$
\begin{eqnarray}\label{4.Q}
Q(y_1,y_2) & := & {1 \over h^3(y_1) h^3(y_2)} \bigg \{
{1 \over \pi} (8y_1^2 y_2^2 - 2y_1^2 - 2y_2^2 + {1 \over 2})
h^2(y_1) h^2(y_2) e^{-2(y_1^2 + y_2^2)} \nonumber \\
&&+ {1 \over \sqrt{2 \pi^3}} \Big (
(12 y_1^2 y_2 - 3y_2) h^2(y_1) h(y_2) e^{2 y_1^2} +
(12 y_1 y_2^2 - 3y_1) h(y_1) h^2(y_2) e^{2y_2^2} \Big ) \nonumber \\
&& \times e^{-4(y_1^2 + y_2^2)}
+ {9 \over \pi^2} y_1 y_2 h(y_1) h(y_2) e^{-4(y_1^2 + y_2^2)} \nonumber \\
&& + {1 \over 2 \pi^2} \Big (
(4y_1^2 - 1) h^2(y_1) e^{-2(y_1^2 + 3 y_2^2)}
+ (4y_2^2 - 1) h^2(y_2) e^{-2(3 y_1^2 + y_2^2)} \nonumber \\
&& + {1 \over \sqrt{8 \pi^5}} \Big ( 6y_1 h(y_1) e^{-2(2y_1^2 + 3y_2^2)}
+ 6y_2 h(y_2) e^{-2(3y_1^2 + 2y_2^2)} \Big ) \nonumber \\
&& + {1 \over 2 \pi^3} e^{-6(y_1^2 + y_2^2)} \bigg \}
\end{eqnarray}
Comparison with (\ref{4.7}) shows that indeed the leading order
behaviour is $\pi^2 \rho_{(1)}(y_1) \rho_{(1)}(y_2)$ so we have
$$
\rho_{(2)}^T(y_1,y_2;x_1 - x_2) \sim {Q(y_1,y_2) \over 4 \pi^2 
(x_1 - x_2)^2}.
$$
Furthermore, a close inspection of (\ref{4.Q}) 
and (\ref{4.7}) reveals that it factorizes as
$$
Q(y_1,y_2) = f(y_1) f(y_2), \qquad f(y) := {d \over dy} \pi \rho_{(1)}(y).
$$
Consequently we have
\begin{equation}\label{4.fin}
\rho_{(2)}^T(y_1,y_2;x_1 - x_2) \sim {f(y_1) f(y_2) \over 4 \pi^2 
(x_1 - x_2)^2}, \qquad \int_{-\infty}^\infty f(y) \, dy = 1,
\end{equation}
which is identical to the behaviour (\ref{4.uf}) for
the OCP at $\beta = 2$, except for the sign.

Note also that since for small separation between $(x_1,y_1)$ and
$(x_2,y_2)$ the truncated distribution is negative, the positive tail for
large $|x_1 - x_2|$ exhibited by (\ref{4.uf}) indicates that $\rho_{(2)}^T$
changes sign, analogous to its behaviour in the bulk. In contrast the same
quantity for the OCP at $\beta = 2$  is always negative, as seen from
(\ref{4.cr4}).

\section{Summary}
The p.d.f.~(\ref{1.3}) when interpreted as a Boltzmann factor consists
of the pairwise logarithmic interaction (\ref{1.5}) as well as an
extensive many body interaction. If the extensive many body interaction
is replaced by a one body Gaussian factor, the p.d.f.~becomes
identical to the p.d.f.~(\ref{1.6}) giving the eigenvalue
distribution for complex Gaussian random matrices, or equivalently
the Boltzmann factor for a 2dOCP. The exact distribution functions
for both (\ref{1.3}) and (\ref{1.6}) can be calculated exactly, enabling
a comparative study of the two systems to be undertaken.

Some similarities, and some differences, have been found. The similarities
include the same $y \to - \infty$ decay of the density profile away from
the boundary being displayed by (\ref{4.8}) for the complex
zeros, and by (\ref{4.8'}) for the soft wall OCP. Also both systems
have the same (up to a minus sign) universal form (\ref{4.uf}) of the
two-point correlation parallel to the boundary. A noteable difference
is that while the second moment of the two-point function in the bulk
of the OCP obeys the Stillinger-Lovett sum rule (\ref{3.lr2}), the same
quantity for the complex zeros vanishes. This was shown to have
consequence regarding the behaviour of the variance of a linear
statistic, a quantity which has received recent attention in the
context of two-dimensional Coulomb systems \cite{17}, and the zeros
of multidimensional random polynomials with a constant density on
the surface of higher dimensional spheres \cite{14}.

\subsection*{Acknowledgements}
This work was initiated by B.~Jancovici who, amongst other things,
found the formula (\ref{3.h1}). The freedom of exchange of ideas,
and the associated stimulating discussions, have been most
appreciated. Also,
we thank P.~Leboeuf for drawing our attention to ref.~\cite{14}.
This work was supported by the Australian Research
Council.


\begin{thebibliography}{99}

\bibitem{1}
P. Leboeuf,
\newblock {\it J. Phys. A}, 24 (1991), 4575

\bibitem{2}
P. Leboeuf and P. Shukla,
\newblock {\it J. Phys. A}, 29 (1996), 4827

\bibitem{3}
P. Shukla,
\newblock {\it J. Phys. A}, 30 (1997), 6313

\bibitem{4}
S. Nonnenmacher and A. Voros,
\newblock {\it J. Stat. Phys.}, 92 (1998), 431

\bibitem{5}
P. Leboeuf,
\newblock Statistical theory of chaotic wavefunctions: a model in terms 
of random analytic functions, preprint

\bibitem{6}
J.H. Hannay,
\newblock The chaotic analytic function, preprint

\bibitem{7}
E. Bogomolny, O. Bohigas and P. Leboeuf, 
\newblock {\it Phys. Rev. Lett.}, 68 (1992), 2726

\bibitem{8}
E. Bogomolny, O. Bohigas and P. Leboeuf, 
\newblock {\it J. Stat. Phys.}, 85 (1996), 639 

\bibitem{9}
J.H. Hannay,
\newblock {\it J. Phys. A}, 29 (1996), L101

\bibitem{10}
T. Prosen,
\newblock {\it J. Phys. A}, 29 (1996), 4417

\bibitem{11}
P. Bleher and X. Di,
\newblock {\it J. Stat. Phys.}, 88 (1997), 269

\bibitem{12}
A. Edelman and E. Kostlan,
\newblock {\it Bull. Amer. Math. Soc.}, 32 (1995), 4365

\bibitem{13}
L.A. Shepp and R.J. Vanderbei,
\newblock {\it Trans. Amer. Math. Soc.}, 347 (1995), 4365

\bibitem{14}
B. Shiffman and S. Zelditch,
\newblock math.CV/9803052

\bibitem{15}
Ph. A. Martin,
\newblock {\it Rev. Mod. Phys.}, 60 (1988), 1075

\bibitem{16}
P.J. Forrester,
\newblock {\it Phys. Reports}, 301 (1998), 235

\bibitem{17}
P.J. Forrester,
\newblock cond-mat/9805306

\bibitem{18}
C. Cohen-Tannoudji, B. Diu and F. Lalo\"e,
\newblock M\'ecanique Quantique (Hermann, Paris, 1977)

\bibitem{18.5}
T.T. Wu and C.N. Yang,
\newblock {\it Nucl. Phys. B}, 107 (1976), 365 

\bibitem{19}
P.A.M. Dirac,
\newblock {\it Proc. R. Soc. London A}, 133 (1931), 60

\bibitem{20}
E. Majorana,
\newblock {\it Nuovo Cimento}, 9 (1932), 43

\bibitem{21}
J. Ginibre,
\newblock {\it J. Math. Phys.}, 6 (1965), 440

\bibitem{22}
B. Jancovici,
\newblock {\it Phys. Rev. Lett.}, 46 (1981), 386

\bibitem{23}
A. Alastuey and Ph. A. Martin,
\newblock {\it J. Stat. Phys.}, 39 (1985), 405

\bibitem{23.1}
A. Erd\'elyi (ed.), 
\newblock Higher Transcendental Functions vol.~1
(McGraw-Hill, New York, 1953)

\bibitem{23.5}
 Ph. A. Martin and T. Yal\c{c}in,
\newblock {\it J. Stat. Phys.}, 22 (1980), 435

\bibitem{24}
B. Jancovici,
\newblock {\it J. Stat. Phys.}, 34 (1984), 803

\bibitem{25}
B. Jancovici,
\newblock {\it J. Stat. Phys.}, 29 (1982), 263

\end{thebibliography}
\end{document}